\documentclass[twocolumn,trackchanges]{aastex701}

\usepackage{subcaption}
\usepackage{amsmath}

\begin{document}

\title{C3NN-SBI: Learning Hierarchies of $N$-Point Statistics from Cosmological Fields with Physics-Informed Neural Networks}

\author[orcid=0009-0007-9746-2074]{Kai Lehman}
\altaffiliation{Equal Contribution; Corresponding authors}
\affiliation{Universit\"ats-Sternwarte M\"unchen, Fakult\"at f\"ur Physik, Ludwig-Maximilians-Universit\"at, Scheinerstr.~1, 81679 M\"unchen, Germany}
\affiliation{Center for Computational Astrophysics, Flatiron Institute, 162 5th Avenue, New York, NY, 10010, USA}
\affiliation{Excellence Cluster ORIGINS, Boltzmannstr.~2, 85748 Garching, Germany}
\email[show]{kai.lehman@physik.lmu.de}  

\author[orcid=0009-0002-7361-4073]{Zhengyangguang Gong} 
\altaffiliation{Equal Contribution; Corresponding authors}
\affiliation{Universit\"ats-Sternwarte M\"unchen, Fakult\"at f\"ur Physik, Ludwig-Maximilians-Universit\"at, Scheinerstr.~1, 81679 M\"unchen, Germany}
\affiliation{Max-Planck-Institut f\"ur extraterrestrische Physik, Giessenbachstr.~1, 85748 Garching, Germany}
\affiliation{Steward Observatory, University of Arizona, 933 North Cherry Avenue, Tucson, AZ 85721, USA}
\email[show]{zgong@arizona.edu}

\author[orcid=0009-0002-2889-3704]{David Gebauer}
\affiliation{Max-Planck-Institut f\"ur extraterrestrische Physik, Giessenbachstr.~1, 85748 Garching, Germany}
\affiliation{Fakult\"at f\"ur Physik, Universit\"at Bielefeld, Postfach 100131, 33501 Bielefeld, Germany}
\email{fakeemail3@google.com}

\author{Stella Seitz}
\affiliation{Universit\"ats-Sternwarte M\"unchen, Fakult\"at f\"ur Physik, Ludwig-Maximilians-Universit\"at, Scheinerstr.~1, 81679 M\"unchen, Germany}
\affiliation{Max-Planck-Institut f\"ur extraterrestrische Physik, Giessenbachstr.~1, 85748 Garching, Germany}
\email{fakeemail4@google.com}

\author[orcid=0000-0002-8282-2010,sname=Asia,gname=Mountain]{Jochen Weller}
\affiliation{Universit\"ats-Sternwarte M\"unchen, Fakult\"at f\"ur Physik, Ludwig-Maximilians-Universit\"at, Scheinerstr.~1, 81679 M\"unchen, Germany}
\affiliation{Excellence Cluster ORIGINS, Boltzmannstr.~2, 85748 Garching, Germany}
\affiliation{Max-Planck-Institut f\"ur extraterrestrische Physik, Giessenbachstr.~1, 85748 Garching, Germany}
\email{fakeemail5@google.com}

\begin{abstract}
Cosmological analyses are moving past the well understood 2-point statistics to extract more information from cosmological fields. A natural step in extending inference pipelines to other summary statistics is to include higher order N-point correlation functions (NPCFs), which are computationally expensive and difficult to model. At the same time it is unclear how many NPCFs one would have to include to reasonably exhaust the cosmological information in the observable fields. An efficient alternative is given by learned and optimized summary statistics, largely driven by overparametrization through neural networks. This, however, largely abandons our physical intuition on the NPCF formalism and information extraction becomes opaque to the practitioner. We design a simulation-based inference pipeline, that not only benefits from the efficiency of machine learned summaries through optimization, but also holds on to the NPCF program. We employ the heavily constrained Cosmological Correlator Convolutional Neural Network (C3NN) which extracts summary statistics that can be directly linked to a given order NPCF. We present an application of our framework to simulated lensing convergence maps and study the information content of our learned summary at various orders in NPCFs for this idealized example. We view our approach as an exciting new avenue for physics-informed simulation-based inference.
\end{abstract}

\keywords{\uat{Astrostatistics techniques}{1886} --- \uat{Astrostatistics tools}{1887} --- \uat{Convolutional neural networks}{1938}  --- \uat{Cosmological parameters from large-scale structure}{340}}

\section{Introduction}

Extracting maximal cosmological information from current~\citep{hetdex_08, kids_13, boss_13, desi_13, sptsz_15, hsc_15, des_16, eboss_16,erosita_21} and upcoming~\citep{lsst_09,euclid_11,4most_12,pfs_14,ska_15,roman_15} large-scale structure surveys increasingly requires moving beyond the standard 2-point correlation function (2PCF, or its Fourier counterpart the power spectrum) framework \citep{krause_25}. Two general strategies have emerged as promising approaches. The first holds on to the use of summary statistics, both analytically motivated such as \cite{Scoccimarro2000Bispectrum, Sefusatti2007Trispectrum, Kratochvil_minkowski_2013, Fluri_peak_counts_2018, Halder_i3PCF_2021, Gong_i3PCF_2023, wang_25}, and machine learning optimized (e.g.~\cite{charnock_18, jeffrey_21, lemos_23})\footnote{The ideas of such compression methods for use in Approximate Bayesian Computation~\citep{rubin_84, beaumont_02, akaret_15} go back to~\cite{blum_08, fearnhead_10, blum_12, jiang_15}}. These approaches aim to compress non-Gaussian information into low-dimensional statistics. Recent comparisons \citep{lanzieri_25} have shown that summary-based methods can capture substantial amount of information beyond the 2-point level, particularly when higher-order statistics are included.
As an alternative to the use of summary statistics, field-level inference has gained attention in which analyses are performed directly on the full density field rather than on compressed summaries \citep{porqueres_23, zhou_24, minh_2024, leclercq_25}. Such approaches can, in principle, access the complete information content of the data, including all higher-order correlation functions encoded in the field.

Both strategies have demonstrated impressive gains. Summary-based approaches can recover large amounts of previously inaccessible information \citep{lanzieri_25}, while field-level methods have shown even larger potential improvements in constraining power \citep{minh_2024}. However, each method comes with significant limitations. Field-level inference typically requires differentiable forward models, repeated simulations, and computationally expensive frameworks with high-dimensional sampling algorithms for the initial conditions, making it a challenging task to scale to realistic survey volumes on extremely high dimensional parameter spaces. Due to the computational expense and the inherent fidelity of such analyses, the use of summary statistics is still of value as a cheaper and more robust alternative. Conversely, optimized summary statistics, especially those produced by neural networks, often suffer from limited interpretability, as the learned compression is a high dimensional non linear function with a large number of learned parameters. Overfitting is another concern, especially given the imperfect nature of cosmological simulations, one would like to produce summary statistics which are not contaminated by learned artifacts of a given simulation.

To address these limitations, \cite{miles_21} introduced the Correlator Convolutional Neural Network (CCNN) architecture in the context of quantum matter, imposing structural constraints on convolution kernels such that each learned summary can be expressed explicitly as a weighted sum of NPCFs at a given order. This yields a transparent connection between network output and analytically interpretable statistics, bridging the gap between interpretability and flexibility.
Building on this idea, \cite{gong_24} adapted and extended the CCNN architecture for cosmological applications, and developed the so called Cosmological Correlator Convolutional Neural Network (C3NN). Their advances included:
\begin{enumerate}
    \item Enforcing rotational and translational invariance appropriate for cosmological density fields, thus encoding the cosmological principles of homogeneity and isotropy into the architecture.
    \item Implementing an auxiliary algorithm (feature selection process) that can hierarchically distinguish contributions from different orders of NPCFs to a given binary classification task.
    \item As a proof of concept, demonstrating that higher-order statistics, particularly bispectrum-related information learned by C3NN, significantly improve the discrimination between Gaussian and lognormal fields as well as simulated weak lensing convergence maps with different underlying equation of state parameter $w_0$ of dark energy, proving that C3NN captures the information following its design.
\end{enumerate} 
With the above features, their work provided the first cosmological toy demonstration that C3NN can isolate higher-order information in a controlled and interpretable way.

In this work, we extend this program from a pure classification task to mock cosmological parameter inference. We integrate a C3NN summary extractor with a simulation-based inference (SBI) framework, and train both components jointly to infer cosmological parameters directly from simulated weak lensing convergence maps. This joint training ensures that the summary extraction is optimized specifically for the downstream inference task.

Our approach provides two key advantages:
\begin{enumerate}
    \item Efficient extraction of higher-order information. Direct computation and measurement of NPCFs beyond the 3-point correlation function (3PCF) is prohibitively expensive, and current cosmological analyses rarely consider orders above $N=3$. The constrained architecture of C3NN enables us to incorporate higher-order information at negligible computational cost while retaining interpretability: each component of the summary can be explicitly associated with a specific $N$-point order.
    \item Quantifying information beyond a given $N$-point order. As we can clearly separate the summary statistic in orders of NPCFs, we can systematically bound the incremental information gained from each additional order.
    This enables us to forecast whether improvements are more efficiently achieved by adding higher NPCF orders or by transitioning to unconstrained machine learning optimized summaries.
\end{enumerate}

The remainder of this paper is organized as follows. Section~\ref{sec:c3nn_methodology} reviews the overall architecture of C3NN-SBI, its convolutional constraints and the explicit correspondence between the model output and NPCFs. Section~\ref{sec:sims_data} describes our mock simulations, targeted cosmological parameter space and training data preparation. Then Section~\ref{sec:results} presents our results on the information content of successive NPCF orders as approximated by our framework and the residual information not captured beyond 4th order. We discuss our validation tests, as well as limitations and caveats in the current framework in Section~\ref{sec:validation_and_caveats}. Finally we present our conclusions and discuss future work in Section~\ref{sec:conclusions}. 

\section{C3NN-SBI model Architecture}
\label{sec:c3nn_methodology}

\subsection{Optimizable N-point Information Extraction with C3NN}
We briefly introduce the C3NN embedding network here. For a more detailed discussion, as well as tests on cosmological toy models in a classification task setting, we refer the readers to \citet{miles_21,gong_24}. The C3NN architecture starts as a standard single-layer Convolutional Neural Network (CNN) without an activation function. Beyond the usual convolutional feature maps, we define so-called \emph{moment maps}, which can be constructed analytically from the output of the first convolutional layer alone.

By taking spatial average of these moment maps, we obtain a set of scalar summary statistics, referred to as \emph{$N$th-order moments}. These moments can be written analytically in terms of the corresponding NPCF estimators of the input fields. The resulting summary statistics therefore admit a clear interpretation as weighted integrals (to be more precise, summation in the case of discrete pixels) over NPCFs, where the weights are determined by the convolutional filters. We optimize these weights by maximizing the variational mutual information between the summary statistics and the inferred cosmological parameters, using standard gradient-based optimization techniques. Further details are provided below.

\subsubsection{Moment Maps and Their Connection to NPCFs}

We define the first operation on an input image $S_k(\pmb{x})$ with pixel coordinates $\pmb{x}$ in the $k$th channel as the first convolutional layer and produce the linear feature map $C^{(1)}_\alpha$ given by
\begin{equation}
C^{(1)}_\alpha(\pmb{x})
=
\sum_{\pmb{a},k}
w_{\alpha,k}(\pmb{a})\,
S_k(\pmb{x}+\pmb{a}),
\end{equation}
where $w_{\alpha,k}(\pmb{a})$ denotes the learnable weight of filter $\alpha$ at offset $\pmb{a}$ from $\pmb{x}$. This operation is the same as the first layer in a traditional convolutional neural network.

In analogy, we define the \emph{$N$th-order moment map} for $N\in \mathbb{N}$ as
\begin{equation}
\begin{aligned}
C^{(N)}_\alpha(\pmb{x})
&=
\frac{1}{N!}
\sum_{\substack{(\pmb{a}_1,k_1),\dots,(\pmb{a}_N,k_N)\\}}^{(\pmb{a}_1,k_1) \neq \ldots \neq (\pmb{a}_N,k_N)}
\prod_{j=1}^{N}
w_{\alpha,k_j}(\pmb{a}_j)\,
S_{k_j}(\pmb{x}+\pmb{a}_j).
\end{aligned}
\end{equation}
where $(\pmb{a}_1,k_1) \neq \ldots \neq (\pmb{a}_N,k_N)$ above the summation does not denote an upper limit, but rather an exclusion of self-counting. We retain such a notation convention below unless otherwise stated.

As a concrete example, the 2nd-order moment map reads
\begin{equation}
\begin{aligned}
C^{(2)}_\alpha(\pmb{x})
&=
\frac{1}{2}
\sum^{(\pmb{a}_1,k_1)\neq(\pmb{a}_2,k_2)}_{\pmb{a}_1,\pmb{a}_2,k_1, k_2}
w_{\alpha,k_1}(\pmb{a}_1)\,
w_{\alpha,k_2}(\pmb{a}_2)\, \\
& \times S_{k_1}(\pmb{x}+\pmb{a}_1)\,
S_{k_2}(\pmb{x}+\pmb{a}_2).
\end{aligned}
\end{equation}

Taking the spatial average of this map yields the corresponding {\it 2nd-order moment},
\begin{equation}
\begin{aligned}
c^{(2)}_\alpha
&=
\frac{1}{N_{\mathrm{pix}}}
\sum_{\pmb{x}}
C^{(2)}_\alpha(\pmb{x}) \\
&=
\frac{1}{2}
\sum^{(\pmb{a}_1,k_1)\neq(\pmb{a}_2,k_2)}_{\pmb{a}_1,\pmb{a}_2,k_1, k_2}
w_{\alpha,k_1}(\pmb{a}_1)\,
w_{\alpha,k_2}(\pmb{a}_2)\, \\
&\times \left[
\frac{1}{N_{\mathrm{pix}}}
\sum_{\pmb{x}}
S_{k_1}(\pmb{x}+\pmb{a}_1)\,
S_{k_2}(\pmb{x}+\pmb{a}_2)
\right].
\end{aligned}
\end{equation}

The term in brackets is the real-space 2-point correlation function (2PCF) estimator between channels $k_1$ and $k_2$,
\begin{equation}
\hat{\xi}_{k_1 k_2}(\pmb{r})
=
\frac{1}{N_{\mathrm{pix}}}
\sum_{\pmb{x}}
S_{k_1}(\pmb{x}+\pmb{a}_1)\,
S_{k_2}(\pmb{x}+\pmb{a}_2) \ ,
\end{equation}
with separation $\pmb{r}=\pmb{a}_2-\pmb{a}_1$. Substituting this definition, the second-order moment can be written compactly as
\begin{equation}
c^{(2)}_\alpha
=
\frac{1}{2}
\sum^{(\pmb{a}_1,k_1)\neq(\pmb{a}_1+\pmb{r},k_2)}_{\pmb{a}_1,\pmb{r},k_1,k_2}
w_{\alpha,k_1}(\pmb{a}_1)\,
w_{\alpha,k_2}(\pmb{a}_1+\pmb{r})\,
\hat{\xi}_{k_1 k_2}(\pmb{r}) \ .
\end{equation}

This construction generalizes straightforwardly to higher orders. For instance, the third-order moment can be expressed as a weighted sum over the corresponding 3PCF estimator $\hat{\zeta}_{k_1 k_2 k_3}$
\begin{equation}
\begin{aligned}
    c^{(3)}_\alpha&=\frac{1}{3!}\sum^{(\pmb{a}_1,k_1)\neq(\pmb{a}_1+\pmb{r},k_2)\neq(\pmb{a}_1+\pmb{r}',k_3)}_{\pmb{a}_1,\pmb{r},\pmb{r}',k_1,k_2,k_3}w_{\alpha,k_1}(\pmb{a}_1)\,w_{\alpha,k_2}(\pmb{a}_1+\pmb{r})\, \\
    &\times w_{\alpha,k_3}(\pmb{a}_1+\pmb{r}')\hat{\zeta}_{k_1 k_2 k_3}(\pmb{r}, \pmb{r}', -\pmb{r}-\pmb{r}') \ .
\end{aligned}
\end{equation}

\subsubsection{Efficient Computation via a Recursion Relation}

As the computational cost of evaluating higher-order moment maps grows rapidly with order, a direct calculation quickly becomes infeasible. To address this, we instead exploit the recursion relation derived in \citet{miles_21}, which allows higher-order moment maps to be computed based on the lower-order maps. The $N$th-order moment map can be expressed as
\begin{equation}
\label{equ:recursion}
\begin{aligned}
C^{(N)}_\alpha (\pmb{x})
&= \frac{1}{N}
\sum_{l=1}^{N} (-1)^{l-1}
\Bigg[
    \sum_{\pmb{a},k}
    w^l_{\alpha,k}(\pmb{a})\,
    S^l_k(\pmb{x}+\pmb{a})
\Bigg] \\
&\times
C^{(N-l)}_\alpha (\pmb{x}) \ ,
\end{aligned}
\end{equation}
where we define $C_{\alpha}^{(0)}(\pmb{x}) = 1$.

This formulation significantly reduces the computational burden from $\mathcal{O}((KP)^N)$ per input map pixel, where $K$ is the number of input data channels and $P$ is the number of pixels in the convolution filter, to $\mathcal{O}(N^2KP)$ and thus enabling efficient scaling to large maps and high moment orders.

\subsubsection{Isotropic Filters and Inductive Biases}
\label{sec:isotropy}

Under standard cosmological assumptions, the underlying fields are statistically isotropic. To improve training efficiency, it is therefore advantageous to encode this known symmetry directly into the machine learning architecture. We achieve this by enforcing isotropy in our C3NN filters, such that all filter pixels at the same radial distance from the central pixel share a common value.

This additional form of weight sharing, beyond that already imposed by the convolutional structure of the network, substantially reduces the number of trainable parameters while explicitly incorporating prior physical knowledge about the data. Our implementation leverages the \texttt{escnn} library\footnote{https://github.com/QUVA-Lab/escnn} \citep{weiler_19, cesa_22}, which provides a principled framework for symmetry-aware neural networks.

In the machine learning literature, architectural constraints such as isotropy are commonly referred to as \emph{inductive biases}. The performance and interpretation of an optimized model are inherently conditioned on these (sometimes implicit) design choices. Consequently, when assessing the information content of cosmological fields within our optimization framework in the following sections, it is important to keep in mind that the results are conditional on the imposed inductive biases. We return to this point explicitly in our discussion of the results.

\subsection{SBI and Loss Function}

We learn summary statistics to be used for SBI~\citep{alsing_19, cranmer_20}. Generally, in this framework, neural networks are trained on simulated data to extract informative, low-dimensional representations of the observations. The choice of loss function in this setting is not unique. Several learning strategies for the extraction of summary statistics have been proposed in the literature. A common option is to train networks using a mean-squared error (MSE) loss with respect to the parameters \citep{fearnhead_10, jiang_15, lemos_23}. Other approaches include so-called moment networks \citep{jeffrey_20}, which are designed to learn informative moments of the marginalized posterior distribution, as well as end-to-end SBI methods that directly optimize a neural density estimator \citep{radev_20, jeffrey_21, lehman_24}. Following the results of \citet{lanzieri_25}, which demonstrate superior performance of the latter strategy, we adopt a full SBI approach in this work.
For a comprehensive introduction to the SBI paradigm, we refer the reader to \citet{cranmer_20}. Applications of SBI in cosmology include, for example, \citet{jeffrey_21, tucci_23, wietersheim_24, lehman_24, gebauer_25, thomsen_25}\footnote{We further refer to https://github.com/smsharma/awesome-neural-sbi for an extensive list of SBI applications.}.

In our setup, the C3NN acts as an \emph{embedding network}, producing a learned summary representation of the data that is passed directly to the SBI inference model. For the density estimation stage, we employ masked autoregressive flows (MAFs) \citep{papamakarios_17} to model the posterior distribution over parameters, an approach commonly referred to as neural posterior estimation (NPE). The corresponding training objective is given by the negative log-posterior,

\begin{equation}
\label{eq:npe_loss}
\begin{aligned}
\mathcal{L}_{\mathrm{NPE}}
&= - \sum_{i}
\log q_\phi\!\left(\theta_i \mid d_i\right),
\end{aligned}
\end{equation}

where $q_\phi$ denotes the neural posterior density estimator with trainable parameters $\phi$. The network is trained on a set of simulated parameter data pairs $\{\theta_i, d_i\}_{i=1}^{N}$ by maximizing the posterior probability assigned to the true parameters.

An alternative SBI strategy is to learn the likelihood function instead of the posterior directly. This approach, known as neural likelihood estimation (NLE), has the advantage that the prior distribution can be modified after training. In this case, the loss function is given by

\begin{equation}
\label{eq:nle_loss}
\begin{aligned}
\mathcal{L}_{\mathrm{NLE}}
&= - \sum_{i}
\log q_\phi\!\left(d_i \mid \theta_i\right).
\end{aligned}
\end{equation}

Once the likelihood has been learned, standard likelihood-based inference techniques, such as Markov Chain Monte Carlo (MCMC), can be used to obtain posterior samples using the approximate likelihood.

\begin{figure*}
    \centering
    \includegraphics[width=\linewidth]{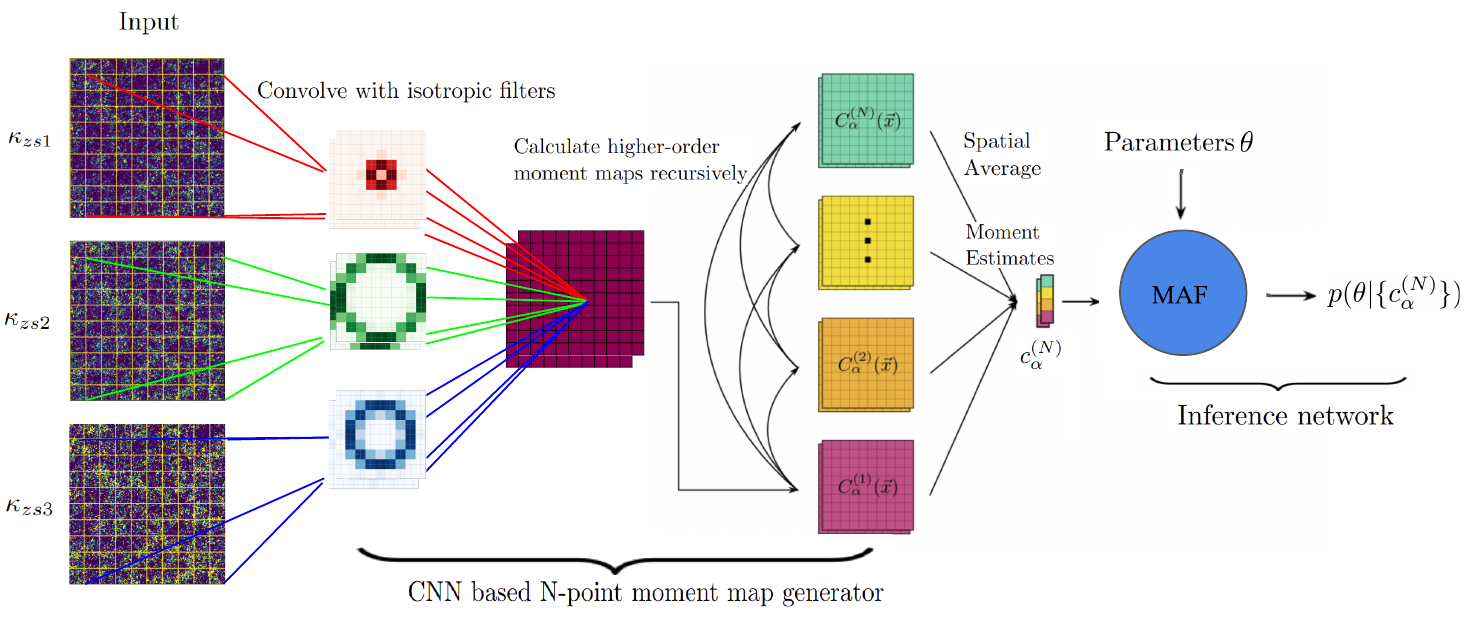}
    \caption{Cartoon illustration of our summary extraction and neural posterior estimation (NPE) pipeline. The input convergence maps ($\kappa$, with $zs1$ to $zs3$ indicating source galaxy tomographic bins, i.e.~different channels) are convolved once with a set of isotropic filters $\alpha$ with learnable weights, producing the first-order moment maps $C_\alpha^{(1)}(\pmb{x})$. Higher-order moment maps $C_\alpha^{(N)}(\pmb{x})$ are then computed recursively using the relation in Eq.~\ref{equ:recursion}, avoiding the need for repeated convolutions. Spatial averaging of each moment map yields the corresponding moment estimates $c_\alpha^{(N)}$, which serve as our summary statistics. These summaries, together with the associated simulation parameters, are passed to a masked autoregressive flow (MAF)–based neural posterior estimator to infer the posterior distribution $p(\theta \mid \{c_\alpha^{(N)}\})$. The entire pipeline is trained end-to-end, with gradients propagating from the posterior estimator through the summary network.}
    \label{fig:cartoon}
\end{figure*}

We present a schematic overview of our C3NN-SBI architecture in Figure~\ref{fig:cartoon}. 
The resulting summaries can be interpreted as a hierarchy of learned, filtered $N$-point statistics, providing a physically motivated and interpretable bridge between traditional moment-based analyses and modern SBI. 
By optimizing these summaries end-to-end for posterior inference, the network learns to combine the information at all orders in NPCFs considered to produce posterior estimates.

\section{Simulations and Data}
\label{sec:sims_data}

In this work, we utilize a suite of simulated weak lensing convergence maps originally generated for the analysis presented in \citet{gebauer_25}. These simulations are designed to emulate the survey properties of the Dark Energy Survey Year 3 (DES Y3) analysis \citep{gatti_21, myles_21} while providing a vast training set suitable for simulation-based inference. We describe the underlying N-body simulations, the lensing methodology, and the data preparation process below.

\subsection{Data \& Lensing}

Weak gravitational lensing provides an unbiased probe of the total matter distribution in the Universe. As light from background sources travels towards the observer, it is deflected by the gravitational potentials of the intervening large-scale structure. In this work, we focus on the weak lensing convergence $\kappa$, a scalar field which describes the isotropic magnification of the source images and corresponds to the weighted line-of-sight projection of the matter density contrast $\delta$. 

Under the Born approximation, assuming a spatially flat Universe, the convergence $\kappa(\pmb{\theta})$ at an angular position $\pmb{\theta}$ on the sky is given by \citep{bartelmann_01}:
\begin{equation}
    \kappa(\pmb{\theta}) = \frac{3 H_0^2 \Omega_m}{2 c^2} \int_0^{\chi_{\mathrm{lim}}} \mathrm{d}\chi \, \frac{\chi}{a(\chi)} g(\chi) \delta(\chi \pmb{\theta}, \chi),
\end{equation}
where $H_0$ is the Hubble constant, $\Omega_m$ is the current matter density parameter, $c$ is the speed of light, and $a(\chi)$ is the scale factor at comoving distance $\chi$. The lensing efficiency kernel $g(\chi)$ depends on the source galaxy redshift distribution $n(z)$ and is defined as \citep{bartelmann_01}:
\begin{equation}
    g(\chi) = \int_{\chi}^{\chi_{\mathrm{lim}}} \mathrm{d}\chi' \, n(\chi') \frac{\chi' - \chi}{\chi'}.
\end{equation}

To generate mock convergence maps from N-body simulations, we rely on the particle-shell approximation \citep{sgier_19, sgier_21, reeves_24}. Instead of continuous integration, the simulation volume is discretized into concentric spherical shells of finite thickness. The effective convergence is then computed as a weighted sum over the particle densities projected onto Healpix maps \citep{gorski_2005} within these shells.

\subsection{The Cosmogrid Sims}

We utilize the CosmoGridV1 dataset \citep{kacprzak_23}, a large suite of N-body simulations designed for simulation-based inference in cosmology. The simulations were evolved using the \textsc{Pkdgrav3} code \citep{potter_17} within periodic cubic boxes of side length $900 \, h^{-1}\mathrm{Mpc}$ containing $832^3$ dark matter particles. The particle mass resolution varies between $3.5 \times 10^{10} \, h^{-1} M_\odot$ and $17.5 \times 10^{10} \, h^{-1} M_\odot$ depending on the specific cosmology.

The grid spans a wide parameter space of flat $w$CDM cosmologies, varying the total matter density $\Omega_m$, baryonic density $\Omega_b$, Hubble constant $H_0$, spectral index $n_s$, fluctuation amplitude $\sigma_8$, and the dark energy equation of state parameter $w_0$. These parameters are sampled using a Sobol sequence to ensure efficient coverage of the high-dimensional space. The suite consists of a wide and a narrow grid with 1250 distinct cosmologies each, with 7 independent realizations per cosmology to suppress cosmic variance. Additionally, 200 independent realizations are provided at a fiducial cosmology to allow for accurate covariance estimation and validation. For the exact distribution of simulation cosmologies in the above parameter space, we would like to refer the reader to Figure~1 in \cite{kacprzak_23}.

The CosmoGridV1 suite incorporates baryonic effects via the baryonification method \citep{schneider_15, giri_21}, applied directly at the map level using particle shells. This method displaces dark matter particles to mimic the reshaping of density profiles due to gas cooling, star formation, and feedback mechanisms. While the full model depends on several parameters, weak lensing observables are primarily sensitive to the mass dependence of the gas profile, parametrized by $M_c$. We adopt the extended parametrization used in the grid:

\begin{equation}
    \label{eq:baryon_feedback}
    M_c(z) = M_c^0 (1+z)^\nu,
\end{equation}

where $M_c^0$ and $\nu$ are varied as free parameters in the simulation grid alongside the cosmological parameters. 

\subsection{Training Data Preparation}
After generating the full-sky simulated weak-lensing convergence maps, we prepare the training data for the C3NN model by partitioning these maps into smaller, square patches. Specifically, we follow the map-partitioning strategy exploited in \cite{gong_24}, which was originally introduced in \cite{ferlito_23}. In this approach, we use each full-sky map to produce non-overlapping square maps of size $50\times50$ pixels.

For each of the simulation nodes in the CosmogridV1 wide grid, we use the same full-sky realization and partition it into 150 non-overlapping square maps following the procedure described above. Each square map contains four channels, corresponding to the four DES Y3 source redshift bins, and it also preserves the angular resolution of the original full-sky simulation, which is generated at NSIDE $=512$. Consequently, the angular size of each square patch is comparable to the largest angular separation considered in the DES Y3 cosmic shear 2-point correlation function analysis \citep{Amon2022, secco_2022}, and the total footprint of the 150 square maps summed up approximately equals to that of DES Y3 footprint. This choice ensures that the spatial extent of the patches remains relevant for direct comparison with scales commonly used in contemporary weak-lensing analyses.

The entire simulations suite consists of 2500 simulations, out of which 1300 are drawn according to a rather wide prior, whereas the remaining 1200 are drawn from a significantly tighter distribution, we refer readers to Figure 1 in \cite{kacprzak_23} for details of the prior distribution in the parameter space. For our main analysis, we restrict ourselves to the simulations stemming from the wide prior (i.e. 1300) but we later investigate the impact increasing the amount of simulations. Furthermore, to construct an independent test data set in our main analysis, we randomly select 250 out of the 1300 cosmological nodes and reserve all of their corresponding square maps exclusively for testing. In addition to the wide-grid simulations, we apply the same partitioning method to the 200 full-sky realizations generated at the fiducial cosmology, where $\Omega_m=0.26$, $\sigma_8=0.84$, $w_0=-1.0$, $n_s=0.9649$, $\Omega_b=0.0493$ and $H_0=67.3$ $\rm{km/s/Mpc}$ \citep{kacprzak_23}. The resulting square maps are subsequently used to produce the results presented in our information content analysis.

The motivation for pushing the pixel size of the square maps to the native angular resolution of the simulations is twofold. First, the simulations already incorporate baryonic feedback effects through the baryonification technique \citep{schneider_15} as described above, making the small-scale information physically meaningful. Second, retaining these small-scale modes allows us to include as much cosmological information as possible in the training data, thereby enabling the C3NN–SBI framework to optimally exploit both large- and small-scale features present in the convergence fields. While observations rely on the estimation of shear from galaxy ellipticities, which introduces shape noise and requires shape measurement calibration, we note that we do not add realistic DES Y3–like shape noise to the training data. The primary goal of this work is not to derive cosmological parameter constraints from observational data, but rather to construct and validate our inference framework and to study the theoretical information content encoded in different orders of correlation functions and in map-level information extractors such as CNNs. Omitting shape noise therefore allows us to isolate the intrinsic information content of the lensing fields themselves. 

\section{Results}
\label{sec:results}

\subsection{Physically Meaningful Hyperparameters and Architectural Choices}

Throughout this work, we only present the best of a few trained networks and defer an exhaustive hyperparameter search to further study. There are, however, some hyperparameters that are of particular importance to our networks as they have clear physical meaning. We motivate the choices of these hyperparameters here. Firstly, the filter size directly determines the largest scale that the network can probe. This is in contrast to traditional CNNs which convolve on the extracted feature map multiple times, effectively also learning scales larger than the initial filter scale. At the same time we want to refrain from setting the filter size too large, as this only adds additional parameters to the network unnecessarily due to the fact that on large scales higher order NPCFs do not contribute significantly to the cosmological information capture. To facilitate symmetries more easily, we further restrict the filter size to be odd in pixel numbers. This provides us with a defined central pixel after each convolution. We can also apply an integer number of padding of zeros to keep the output $C_{\alpha}^{(1)}(\pmb{x})$ maps the same size as the input maps. The filter size chosen for the models presented here is 37, which corresponds to a maximum scale of $r_\mathrm{max} \approx 248'$. This scale is approximately equivalent to the largest angular separation probed by the DES Y3 cosmic shear analyses \citep{Amon2022, secco_2022}. A second important hyperparameter in our models is the number of filters. The amount of scales and configurations that a single filter can reasonably learn is finite. The amount of filters therefore gives an upper limit on how much information can be extracted. Our strategy for this hyperparameter consists of increasing the filter number at a given order in moments, until the constraining power no longer improves. We have found empirically that the 2nd order C3NN embedding network already converged with only a few filters, as opposed to higher order models. This is consistent with our intuition that the possible amount of configurations drastically increases beyond the second order in NPCFs. For models at 2nd order we use 4 filters, at 3rd order 25, and at 4th order we use 40 filters. Increasing the number of filters at any order beyond these numbers has not resulted in any significant improvement. 

As our data consists of 150 maps that add up to the size of DES Y3 footprint, our architecture computes the moments for each of these maps. We then take the mean of the moments as the summary statistic. We furthermore found it very effective, to not only use the means, but also the standard deviation of the moments. This is because in principle, the standard deviation of a given order $N$ moment can already access part of the information contained in the connected 2$N$-point correlation function, though the square root operation would ensure that the standard deviation has the same physical units and scaling behavior as the moment itself. The final length of our summary statistic vector, the set of all moment estimates $\{c_\alpha^{(N)}\}$ is therefore given by:

\begin{equation} \label{equ:dimensionality}
    \mathrm{dim}(\{{c_\alpha^{(N)}}\}) = 2 \times N_\mathrm{filter} \times N.
\end{equation}

As we always use all filters at a given order, we only denote the $N$-point order as a subscript when showing our results in the following.

\subsection{Information Gain at Different Orders}
\label{sec:info_gain_c2_c3_c4}

\begin{figure*}
     \centering
     \begin{subfigure}[b]{.49\textwidth}
         \centering
        \includegraphics[width=\textwidth]{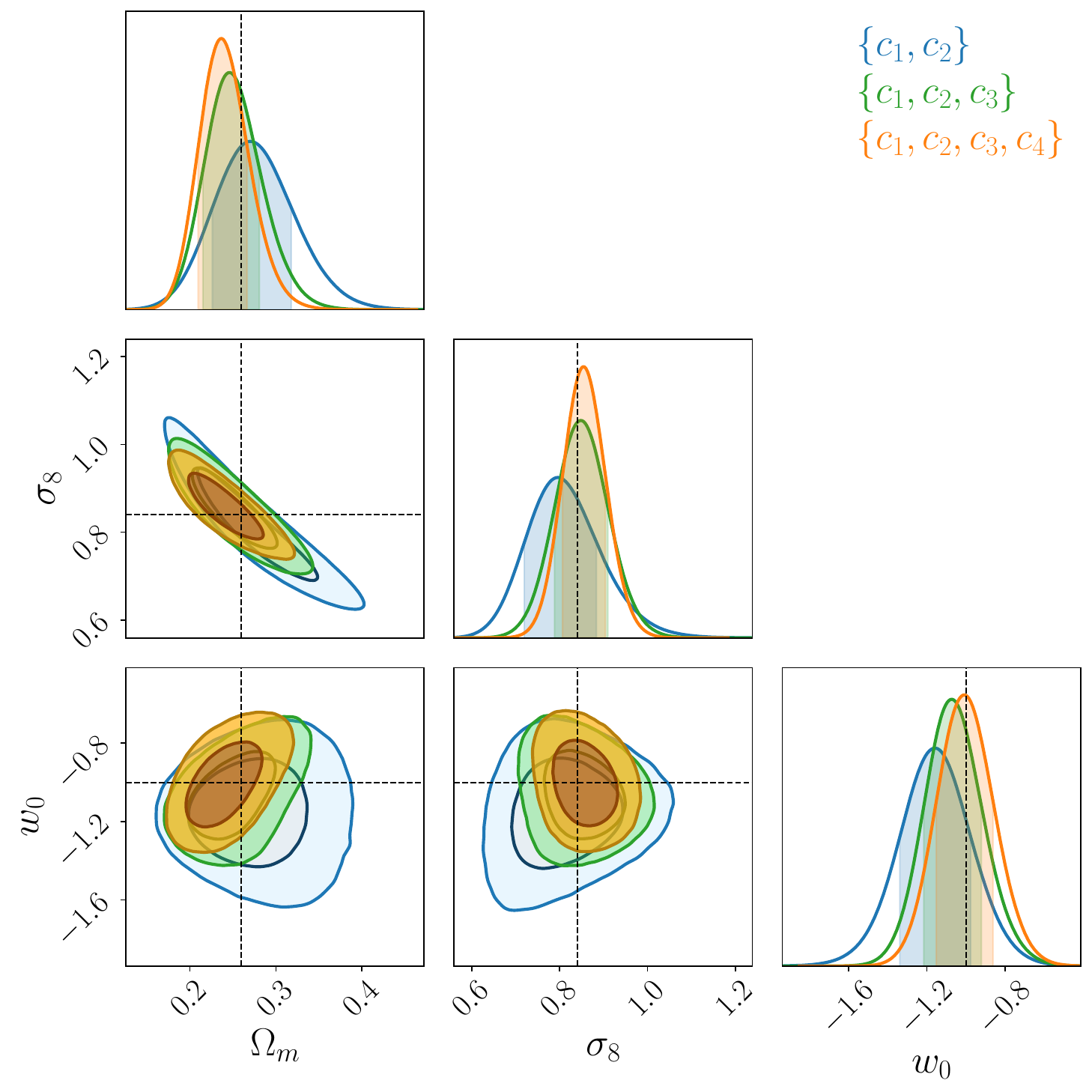}
        \caption{}
        \label{fig:C3NN_orders}
     \end{subfigure}
     \hfill
     \begin{subfigure}[b]{.49\textwidth}
         \centering
        \includegraphics[width=\textwidth]{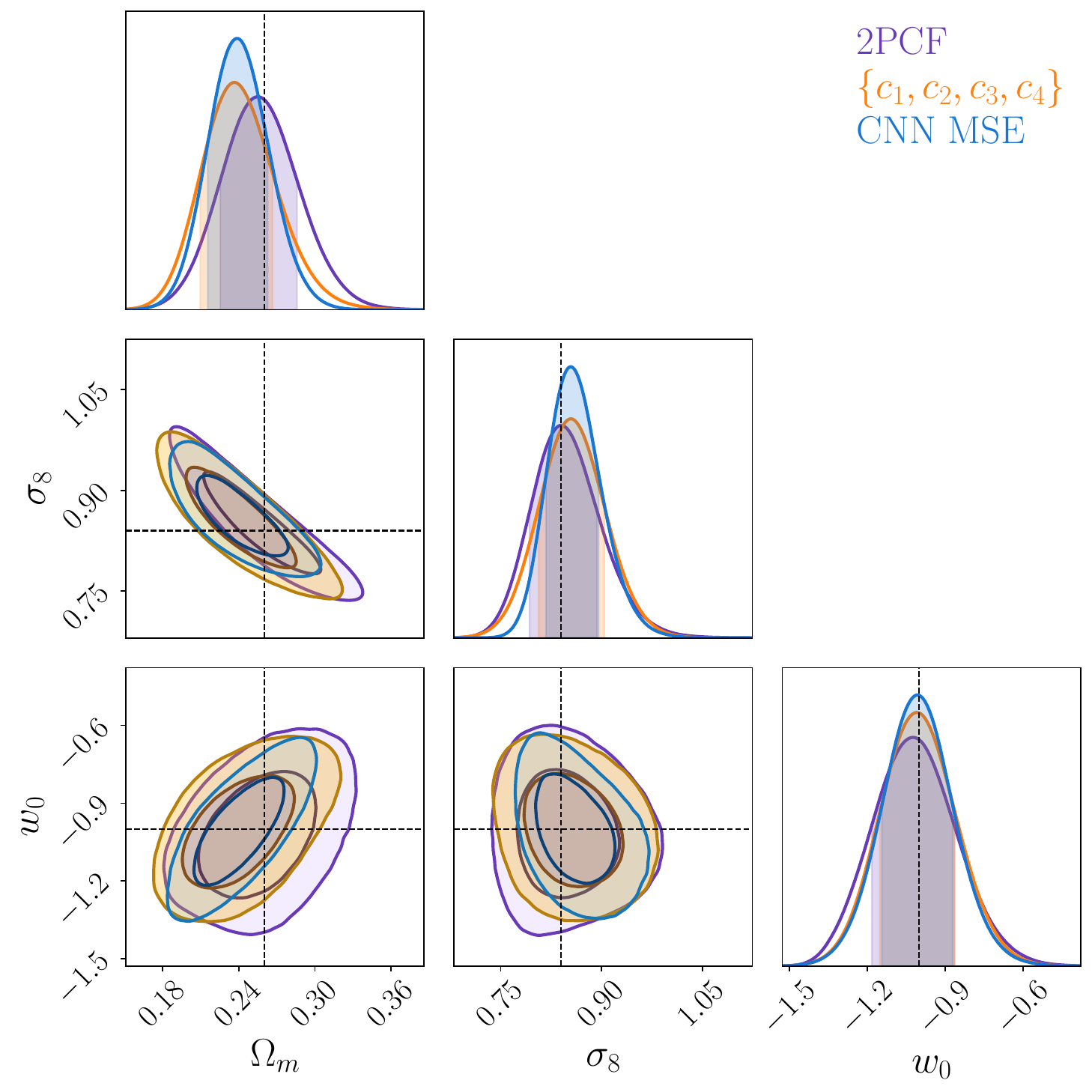}
        \caption{}
        \label{fig:CNN}
     \end{subfigure}
    \caption{Mock inference on a withheld test simulation at the fiducial cosmology. In (a) we compare the constraining power of C3NNs at different orders. As we increase the order, i.e.~up to the 4th order which is the highest order of correlation function considered, the constraining power increases due to the non-gaussian information in the convergence map. In (b) we compare a traditional CNN with our C3NN at 4th order. We find that at the same low amount of training simulations, the CNN model can extract more cosmological information possibly drawing from even higher orders in NPCFs. We also add the mock inference results from measured 2PCFs, and its constraining power is comparable to our C3NN at 4th order, due to the limited training simulation budget.}
    \label{fig:chis}
\end{figure*}

In Figure~\ref{fig:C3NN_orders} we show the resulting contours of C3NNs trained at different orders for mock inference on a previously withheld simulation. Generally, we find that we can overcome the prior in $\Omega_m$, $\sigma_8$ and $w_0$. As we have found in preliminary tests that the other parameters are not constrainable within the prior, we do not explicitly infer them here and instead implicitly marginalize over them, including $\Omega_b$, $H_0$, $n_s$, and $M_c^{0}$, $\nu$ controlling the baryonic effects (Eq.~\ref{eq:baryon_feedback}). In the constrainable parameters we encouragingly find a monotone increase in constraining power as we move to higher orders of moment maps, i.e. to higher order in NPCFs. We did not find any improvement in constraining power after the 4th order. However, we suspect this to be an artifact of the overall low training simulation budget at our disposal and we will investigate this further in the following sections. Concerning inference at 5th order, which in theory should be achievable within our framework, we defer such investigations to future work.  

We provide in Table~\ref{tab:C3NN_c2_c3_c4} the relative improvement in terms of the fractional difference on the marginalized standard deviation of each of the three parameters as well as the 3-dimensional Figure of Merit (FoM) relative improvement when we move to higher orders. All the numbers presented in the table are acquired by running the models on a single fiducial cosmology realization.

\begin{table}[h]
\centering
\begin{tabular}{c c c c c}
\hline
Model & $\Omega_m$ & $\sigma_8$ & $w_0$ & FoM \\ \hline
$\{c_1, c_2, c_3\}$ & $27.92\%$ & $28.44\%$ & $19.11\%$ & $96.59\%$ \\
$\{c_1, c_2, c_3, c_4\}$ & $12.89\%$ & $19.99\%$ & $3.68\%$ & $34.38\%$ \\ \hline
\end{tabular}
\caption{The relative improvement in terms of the fractional difference of the 1D marginalized parameter standard deviation and the FoM from the 3D parameter covariance. The percentage numbers of model $\{c_1, c_2, c_3\}$ are with respect to model $\{c_1, c_2\}$, and those of model $\{c_1, c_2, c_3, c_4\}$ are with respect to model $\{c_1, c_2, c_3\}$.}
\label{tab:C3NN_c2_c3_c4}
\end{table}

\begin{figure*}
     \centering
     \begin{subfigure}[b]{.49\textwidth}
         \centering
        \includegraphics[width=\textwidth,height=0.3\textheight]{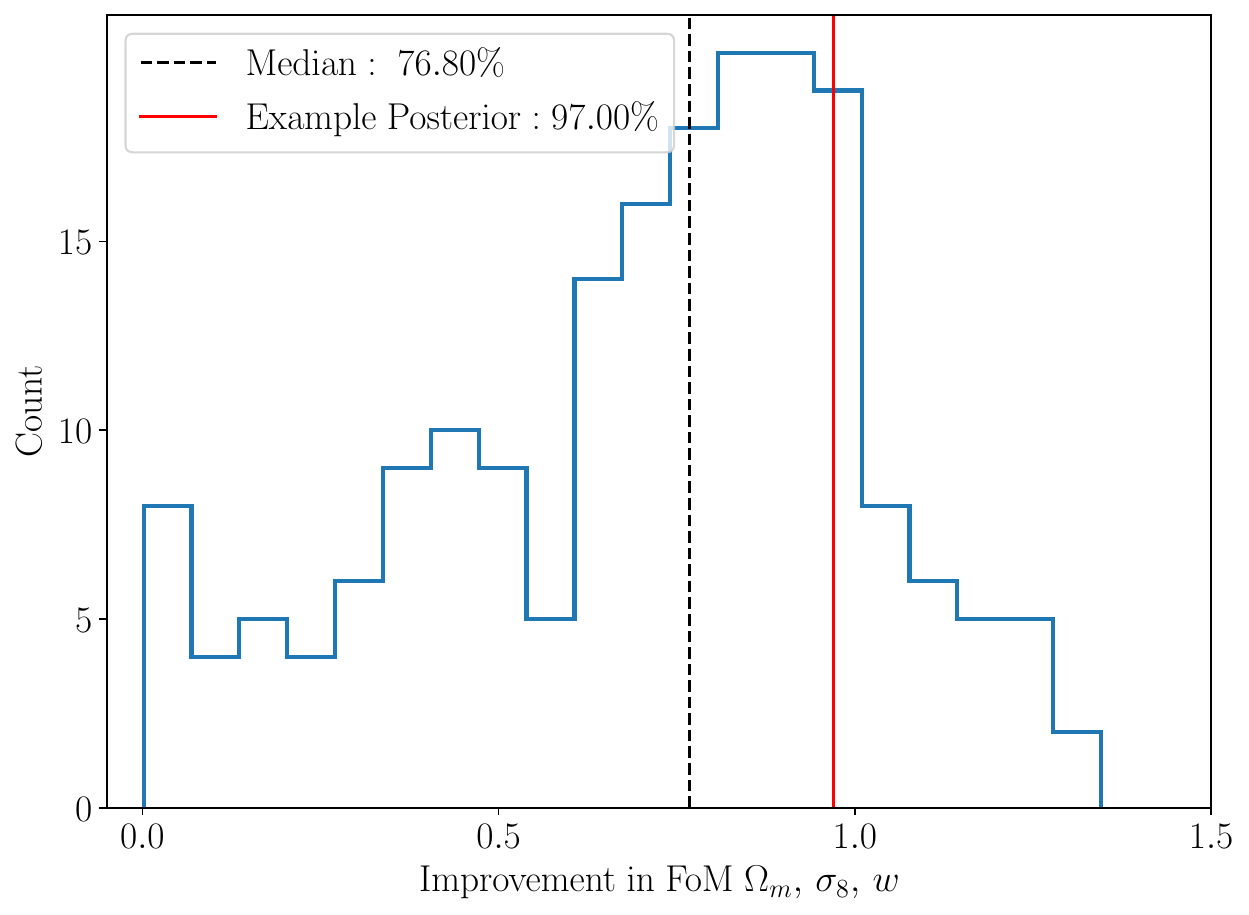}
        \caption{}
        \label{fig:FoM_c3_c2}
     \end{subfigure}
     \hfill
     \begin{subfigure}[b]{.49\textwidth}
         \centering
        \includegraphics[width=\textwidth,height=0.3\textheight]{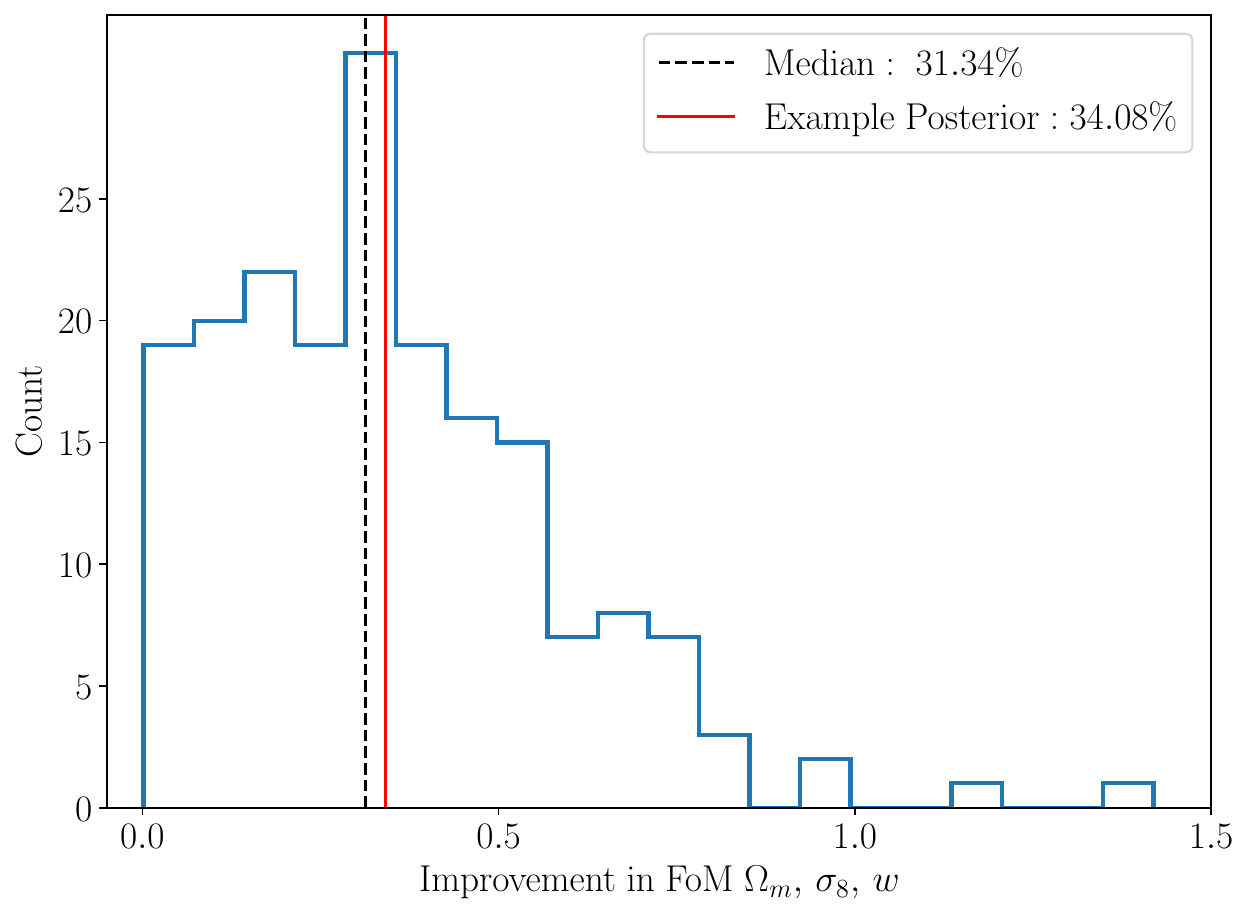}
        \caption{}
        \label{fig:FoM_c4_c3}
     \end{subfigure}
    \caption{Histograms of the improvement in the FoM in terms of fractional difference for 200 simulation realizations at the fiducial cosmology. We also show the median improvement in black, and the improvement seen in the posterior of the selected fiducial simulation inference shown in Figure~\ref{fig:C3NN_orders} in red.}
    \label{fig:FoM_dist_c2_c3_c4}
\end{figure*}

We further run the same test as the one shown in Table~\ref{tab:C3NN_c2_c3_c4}, but on 200 independent simulation realizations on fiducial cosmology.  Figure~\ref{fig:FoM_dist_c2_c3_c4} shows histograms of the improvements in the FoM in terms of fractional difference from adding the $c_3$ statistic on top of $c_2$ (Figure~\ref{fig:FoM_c3_c2}) and adding the $c_4$ statistic on top of $c_3$ (Figure~\ref{fig:FoM_c4_c3}). While the improvement varies for different realizations, the improvements for the posterior shown in Figure~\ref{fig:C3NN_orders} are rather characteristic of and close to the median improvement, and thus are representative of the improvement one would get from adding higher order $N$-point information using the C3NN model. The analysis demonstrates a substantial and consistently positive impact from including the higher-order $N$-point information, with a median FoM improvement of $75.30\%$ from adding $c_3$, and a median FoM improvement of $30.69\%$ from adding $c_4$. 

\subsection{Comparison to a Traditional CNN}

We further compare the information that can be extracted with a standard CNN, in order to assess the quality of information extraction from the models presented thus far while drawing the comparison at the same amount of training budget. It is important to note that this test is not run to assess if all information is extracted, as the CNN itself may fail to do so. The CNN model implemented in this section consists of 3 convolutional layers with ReLU activation function in between, followed by one max pooling layer after each nonlinear activation. We then take the average over the output realizations and apply an adaptive pooling layer. This result is then fed into a fully connected 3-layer neural network with ReLU between it.

Our implementation of C3NN at 4th order only has about 3000 trainable parameters, and the CNN model approximately has 165000 parameters. For this test we train the CNN on a mean square error loss regressing to the standardized cosmological parameters. This results in an estimate of the posterior mean~\cite[see e.g.][]{murphy_22}, which is then fed as a summary statistic into an NPE network. The latter is trained in a separate process. In Figure~\ref{fig:CNN} we show the comparison of the contours obtained from the CNN model above with such an optimization strategy against our best C3NN model in 4th order, when trained on the same low number of simulations. We find that the CNN model has a better performance on the interested cosmological parameters than the C3NN at 4th order, even with this limited number of training simulations. We identify two possible contributions to this difference:

\begin{enumerate}
    \item The CNN has access to higher-order information that is not probed by a chosen finite set of correlation functions in the C3NN case. 
    \item On the orders of NPCFs that the two models do share, C3NN is limited to learning weighted integrals of NPCFs while the CNN has more freedom. This freedom comes at the cost of opaque information extraction, whereas C3NN retains a clear analytical connection to the NPCF program.
\end{enumerate}

Nevertheless, we remind the readers that these results are subject to the limited training data used. While we find the difference in constraining power intriguing as to the uncaptured information by C3NN in the field, this is merely a hint and could be driven by the impact of the limited amount of training simulations. Thus the robust quantitative investigation of the amount of cosmological information remained uncaptured by C3NN to a given correlation function order, but existent in the field, is left to future studies.

\subsection{Comparison to the 2PCF and Hybrid Summaries}

Given that higher-order statistics aim to outperform the 2PCF, we explicitly compare the two in this section. For this we first measure the convergence 2PCFs on the same 1250 wide grid cosmology projected square maps used for training C3NN in 10 logarithmically spaced bins between 10 and 250 arcminutes. We train NPE with a linear, one layer deep, feed forward embedding network on the measured 2PCFs directly from simulation maps. Figure~\ref{fig:CNN} shows the comparison of this posterior with our best C3NN model at 4th order (here the 2PCF model exploits the same amount of training simulations as the C3NN, i.e.~1000 nodes). Surprisingly, the 4th order C3NN model only adds little information on top of the 2PCFs. We trace this back to the limiting small simulation budget available here. This is formally explained by the Dodelson-Schneider correction~\citep{dodelson_13} which is also present in neural posterior analyses~\citep{homer_25}.  We assess two strategies to still outperform the 2PCFs. One is that of hybrid, 2PCF aware, summary statistics\citep{makinen_24,makinen_25}, the other is an increase in training budget by switching to NLE which we perform in the following section. 

\begin{figure*}
    \centering
    \begin{subfigure}[b]{.49\textwidth}
    \centering
    \includegraphics[width=\linewidth]{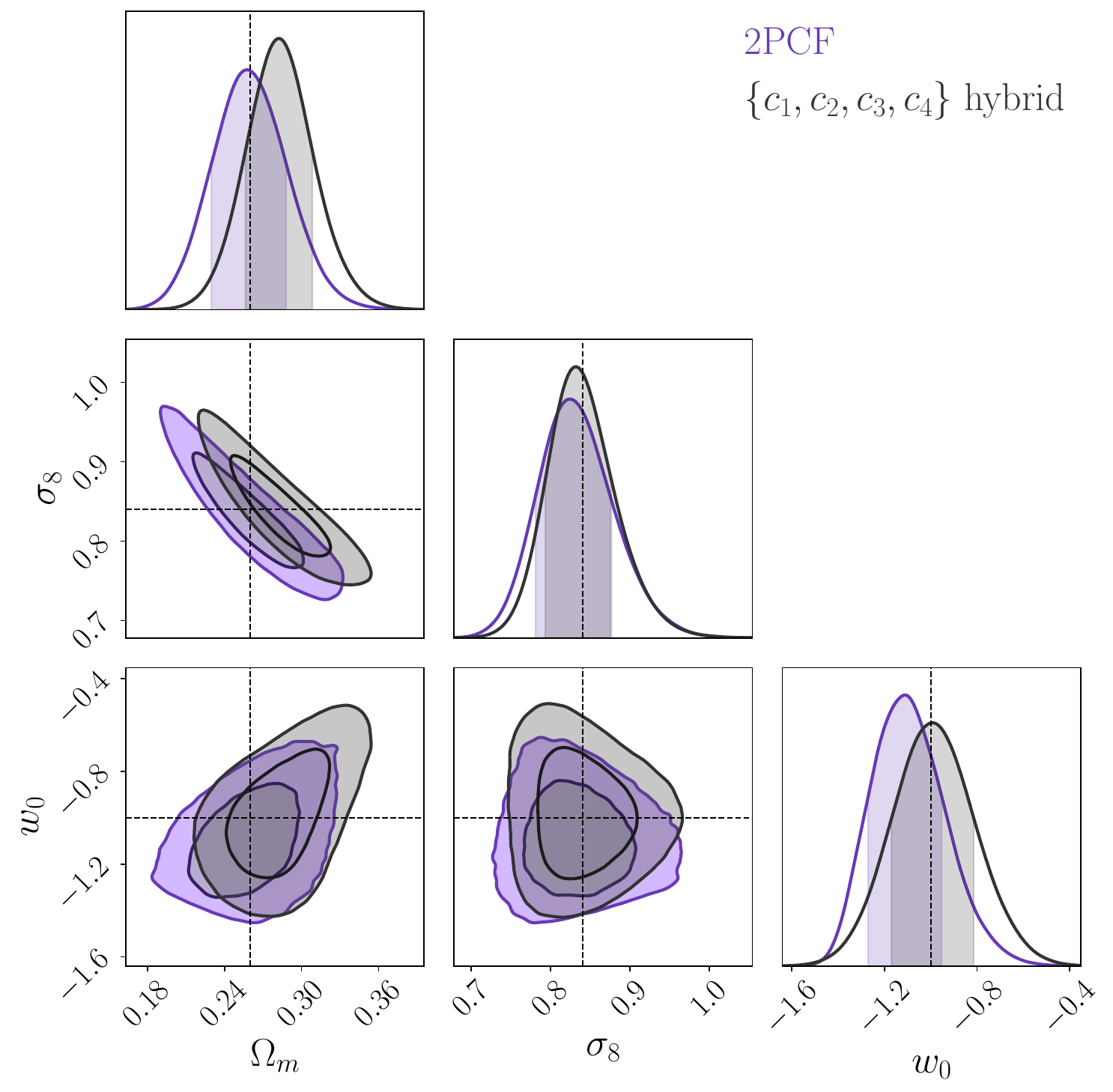}
    \caption{}
    \label{fig:hybrid}
    \end{subfigure}
    \hfill
    \begin{subfigure}[b]{.49\textwidth}
    \centering
    \includegraphics[width=\linewidth]{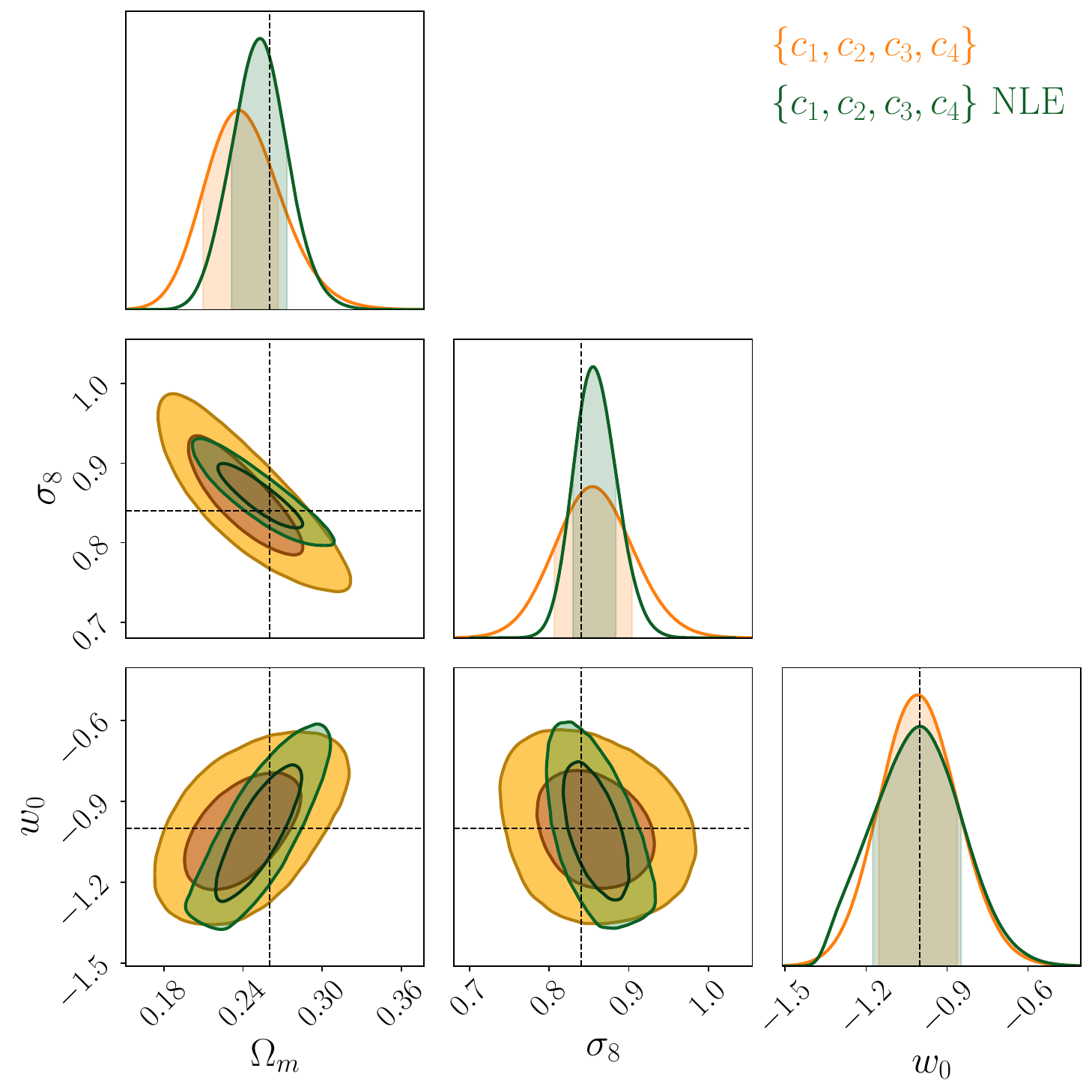}
    \caption{}
    \label{fig:nle}
    \end{subfigure}
    \caption{(a) Comparison of a hybrid~\citep{makinen_24,makinen_25} C3NN at 4th order to the measured 2PCF with NPE. The hybrid model outperforms the 2PCF NPE posterior. (b) Comparison of our best C3NN 4th order NPE result with a corresponding C3NN 4th order NLE model using twice the training data for the inference network. By including more simulations in the training of the NLE inference network, the constraining power increases substantially.}
\end{figure*}

In order to facilitate easier information extraction, based on the knowledge of the 2PCF we first investigate a hybrid approach, as introduced by~\cite{makinen_24,makinen_25}. They have shown, that if the density estimator is not only conditioned on the learned summary, but also an already known summaries (such as the 2PCF estimator in this case), the learned summary improves in information content as the machine can ``concentrate" solely on information beyond the known summary. We train our 4th order C3NN model as a hybrid model conditioned also on the NPE compressed 2PCFs and show the results in Figure~\ref{fig:hybrid}. In this framework the C3NN can learn additional information and leads to the outperformance of the 2PCFs (Tab.~\ref{tab:nle}). As with our other results thus far, we conjecture that the amount of information gained is also still limited by the small amount of available simulations.

\subsection{Increasing the Simulation Budget through NLE}

We can also use the trained C3NN summary network to retrain an inference network based on NLE. In that case, we can include more simulations from CosmoGridV1 suite, as the training simulations are no longer bound to follow the prior distribution. We present the NLE alternative to the NPE posteriors here, in order to investigate whether our pipeline thus far is really simulation limited as conjectured. NLE presents a few downsides in our use case. Most notably, the previously learned summary statistic is no longer optimal under the prior, as the training distribution for the C3NN embedding net is now different from the actual one. In our particular case, the learned summary is also quite high dimensional (see Eq.~\ref{equ:dimensionality}). As the density estimation task is now data dimensional, as opposed to parameter dimensional, this would introduce the curse of dimensionality. In order to make the density estimation feasible, we train NLE on the principal components of the learned summary, and only keep 20 of them. We emphasize that this additional linear  compression introduces {\it heavy losses in information}. It does, however, enable us to now investigate the dependency of our machinery on the simulation budget. 

As this subsection only constitutes such a test, we accept the highly non-optimal postprocessing of the learned summary and our mock posteriors are to be interpreted as {\it lower bounds} on the actual information content of the summary. We present our NLE posteriors with twice the amount of training data in Figure~\ref{fig:nle}. We find that the increase in constraining power is significant, just from the increased simulation budget. The tightened NLE contours at 4th order easily outperform the 2PCF. We report detailed improvements in Table~\ref{tab:nle}. These results confirm our conjecture that we are in fact simulation limited in our particular case.

\begin{table}[h]
\centering
\begin{tabular}{c c c c c}
\hline
Model & $\Omega_m$ & $\sigma_8$ & $w_0$ & FoM \\ \hline
Hybrid vs. 2PCF & $ 9.46\%$ & $  10.36 \%$ & $-10.79\%$ & $7.29\%$ \\
NLE vs. NPE & $36.67\%$ & $79.70\%$ & $-9.41\%$ & $347.73\%$ \\
\end{tabular}
\caption{Same as Table~\ref{tab:C3NN_c2_c3_c4} but comparing (a) the hybrid trained C3NN model at 4th order with the 2PCF at the same simulation budget and (b) the NLE result at 4th order using double the simulation budget with respect to the NPE result at 4th order.}
\label{tab:nle}
\end{table}

\section{Validation}
\label{sec:validation_and_caveats}

In this section we validate our pipeline from two different perspectives. On one hand, given that our NPE posteriors do not significantly outperform the 2 point function (even though the NLE posteriors do), we construct a test to ensure that we are in fact picking up on higher order information. We do so by artificially removing higher order information and comparing the remaining constraining power of our pipeline. On the other hand, we run standard Bayesian inference tests and validate the resulting marginal posteriors with the popular simulation-based calibration (sbc) test~\citep{Talts_SBC_2018}.

\subsection{Phase Reshuffling}

\begin{figure}
    \centering
    \includegraphics[width=\linewidth]{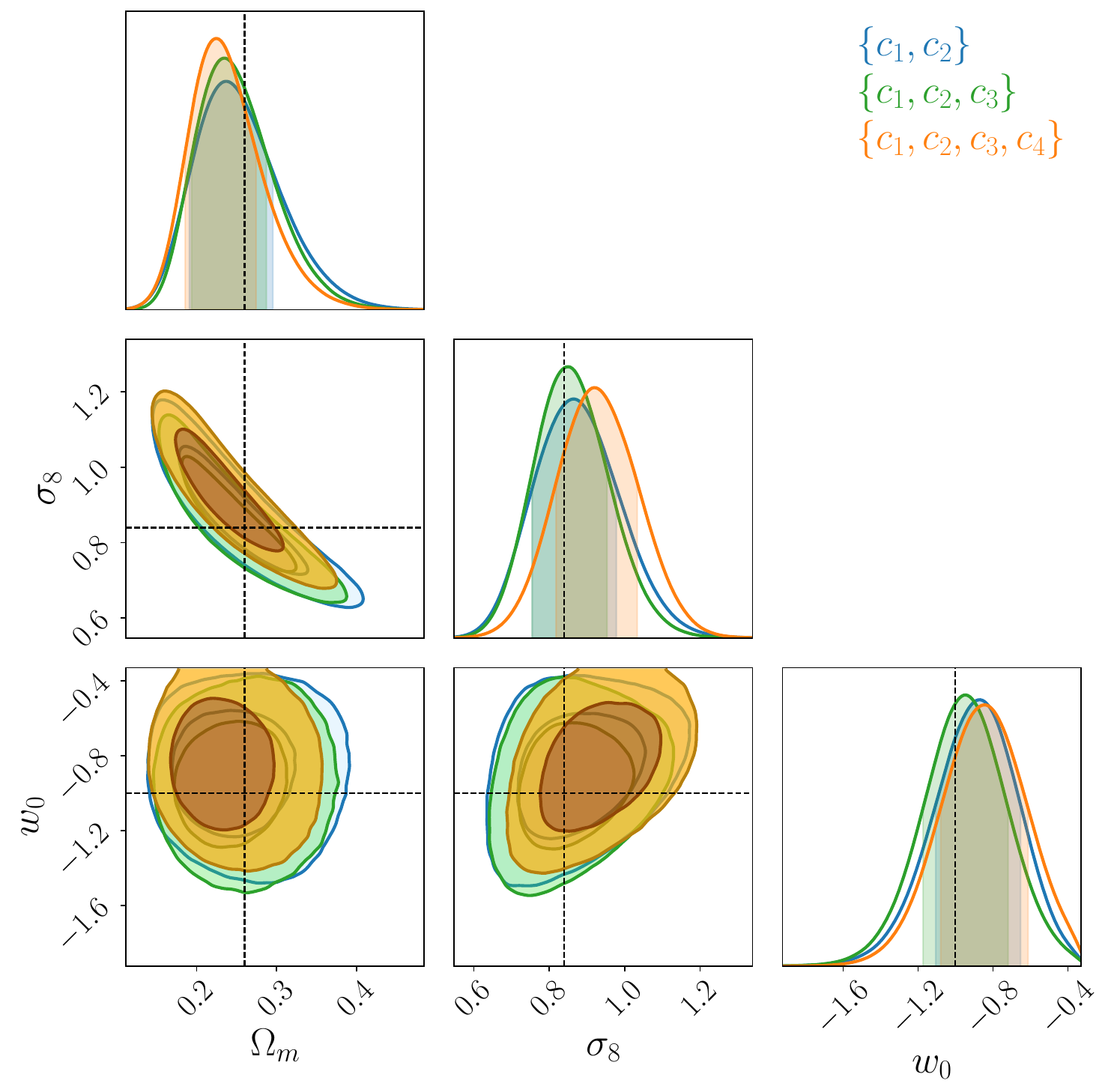}
    \caption{The constraining power of C3NN at different orders where the training maps are on the same cosmological nodes as used in Figure~\ref{fig:C3NN_orders} but {\it phase-randomized}: The fields are first Fast Fourier Transformed (FFT), then their phases randomized following a uniform distribution within 0 and $2\pi$, and then with inverse FFT transformed back to configuration space.}
    \label{fig:C3NN_orders_random_phase}
\end{figure}

Furthermore, we perform a test focusing on phase information in simulated weak lensing convergence maps. We conduct a controlled test where the phase components of our original training data maps are randomized while preserving their amplitude, and then re-evaluate the performance of our C3NN models on these phase-randomized datasets. Figure~\ref{fig:C3NN_orders_random_phase} shows that $c_2$ (2PCF/power spectrum) retains nearly all of its cosmological information content relative to the original training case. In contrast, the information gain derived from $c_3$ (3PCF/bispectrum) and $c_4$ (4PCF/trispectrum) is diminished by a large amount (on certain marginalized parameter dimension there is even information loss) when comparing the results between Table~\ref{tab:C3NN_c2_c3_c4} and~\ref{tab:C3NN_c2_c3_c4_random_phase}. This outcome aligns with theoretical predictions for random Gaussian and non-Gaussian fields: the power spectrum, as a second-order statistic, only encodes amplitude-dependent information about density fluctuations and is thus invariant to phase permutations, while the bispectrum and trispectrum, as higher-order correlation functions, are inherently sensitive to the phase coherence of the field. These phase-sensitive statistics capture the non-random, physically meaningful alignments of structure (e.g. filamentary connections or halo clustering) that encode unique cosmological signatures. Their degradation under phase randomization confirms that our C3NN models are indeed leveraging higher-order phase information to constrain cosmological parameters beyond the limits of power spectrum. 

\begin{table}[h]
\centering
\begin{tabular}{c c c c c}
\hline
Model & $\Omega_m$ & $\sigma_8$ & $w_0$ & FoM \\ \hline
$\{c_1, c_2, c_3\}$ & $7.70\%$ & $14.21\%$ & $-3.93\%$ & $34.89\%$ \\
$\{c_1, c_2, c_3, c_4\}$ & $6.28\%$ & $-4.90\%$ & $2.63\%$ & $-8.23\%$ \\ \hline
\end{tabular}
\caption{The relative improvement of the 1D marginalized parameter standard deviation and the FoM from the 3D parameter covariance. Here the training data have the same size but {\it are randomized on their phases}. The percentage numbers of model $\{c_1, c_2, c_3\}$ are with respect to model $\{c_1, c_2\}$, and those of model $\{c_1, c_2, c_3, c_4\}$ are with respect to model $\{c_1, c_2, c_3\}$. All models trained here possess the same hyperparameters as the corresponding ones previously.}
\label{tab:C3NN_c2_c3_c4_random_phase}
\end{table}

\subsection{Simulation-Based Calibration (SBC) test}

\begin{figure*}[h!] 
    \centering  
    \begin{subfigure}[b]{0.49\textwidth}
        \centering
        \includegraphics[width=\linewidth]{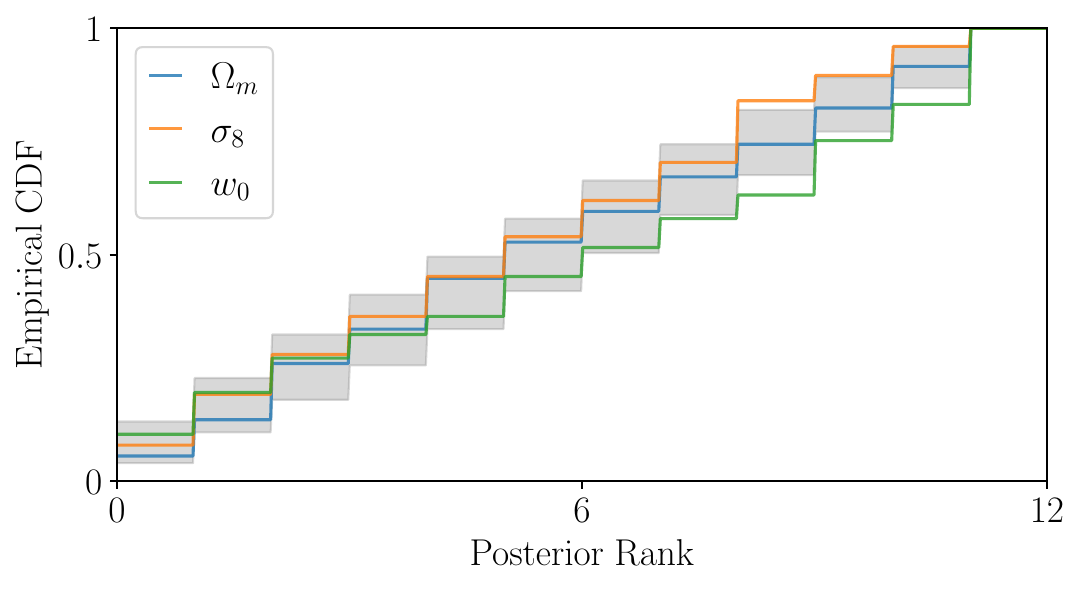} 
        \caption{}
        \label{fig:sbc_c3nn_sub1}  
    \end{subfigure}
    \hfill  
    \begin{subfigure}[b]{0.49\textwidth}
        \centering
        \includegraphics[width=\linewidth]{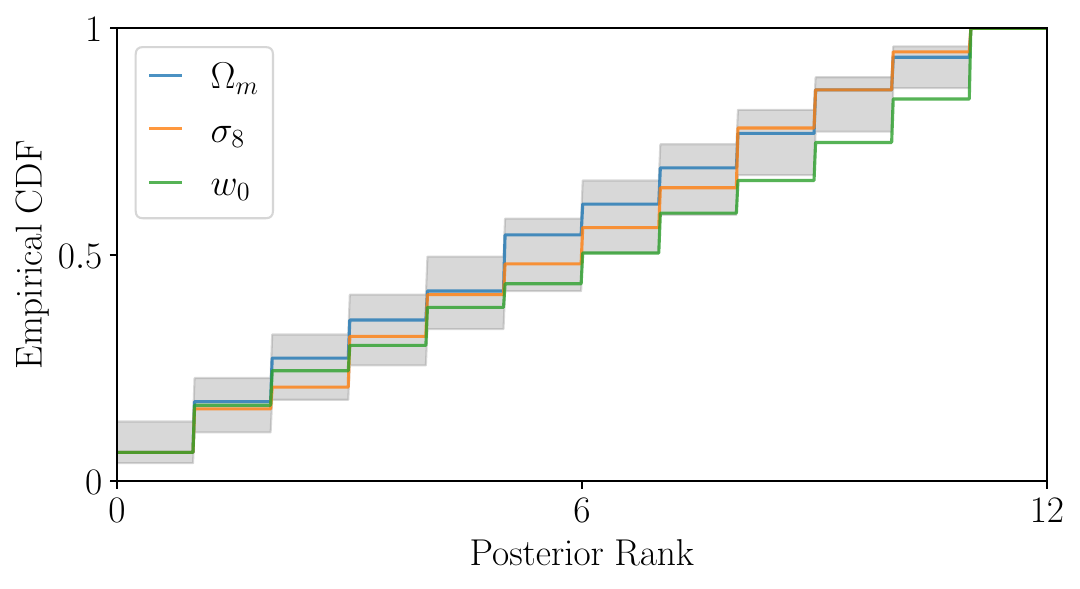}  
        \caption{}
        \label{fig:sbc_c3nn_sub2}  
    \end{subfigure}
    \vspace{0.5cm}
    \begin{subfigure}[b]{\textwidth}  
        \centering
        \includegraphics[width=0.49\textwidth]{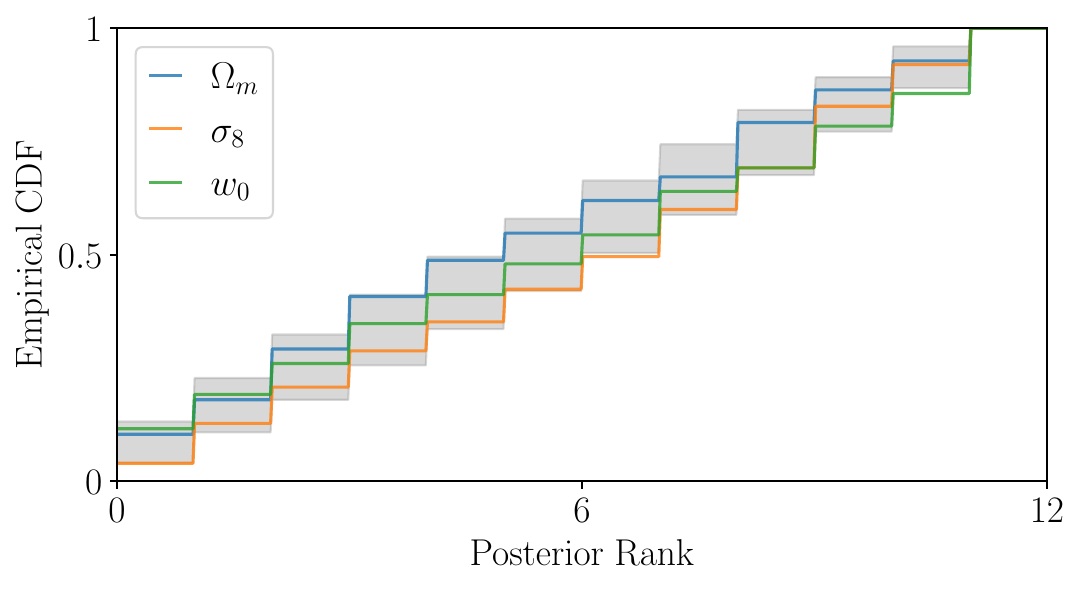}  
        \caption{}
        \label{fig:sbc_c3nn_sub3}  
    \end{subfigure}
    \caption{SBC test for the three C3NN models on a separate test dataset composed of 250 randomly sampled simulations from the CosmogridV1 wide grid for the (a) second order (b) third order and (c) fourth order model. Blue, orange and green curves denote empirical CDF curves of rank statistics for $\Omega_m, \sigma_8$ and $w_0$ respectively. The gray bands represent the overall statistical (95\%) uncertainties of the estimated empirical CDF.}
    \label{fig:sbc_c3nn}  
\end{figure*}

We show the SBC test results for the three models in Section~\ref{sec:info_gain_c2_c3_c4} with different orders of correlators in Figure~\ref{fig:sbc_c3nn}. The test data set is initially separated from the training data set in the CosmogridV1 simulation wide grid. 

SBC is a validation method for Bayesian Inference methods, designed to assess whether a model’s predicted posterior distributions are statistically consistent~\citep{Talts_SBC_2018}. SBC makes use of a so called rank statistic which is used to compare the (test) data averaged marginal posterior to the prior. For a well-calibrated model, these rank statistics should be uniformly distributed over the interval $[0, 1]$. Instead of showing these histograms directly, we show the cumulative distribution function (CDF) which makes it easier to spot deviations.

The behavior of the CDF of the rank statistics provides direct insight into the properties of the trained model: (i) If the posterior distributions are overly broad, the rank CDF would first lie below the identity line, then cross, and lie above for large ranks. This indicates that the model is underconfident. (ii) If the posterior distributions are too narrow, the corresponding CDF curves yield the opposite pattern compared to the underconfident case, implying that the model is overconfident, and (iii) a biased posterior shifts the rank distribution toward one side, which means the CDF curves will lie systematically either below or above the identity line. 

From the SBC results shown in Figure~\ref{fig:sbc_c3nn} for the three cosmological parameters of interest, $\Omega_m, \sigma_8$ and $w_0$, we find that the empirical CDFs of the rank statistics for all three independently trained C3NN models closely follow the identity function and the histograms themselves therefore are uniform. This demonstrates that, when validated on the held-out test dataset, the models exhibit no signs of underconfidence, overconfidence, or systematic bias on the level of their marginal posteriors. 

\subsection{Caveats Concerning Information Content}

We would like to clearly point out the caveats of our analysis. Most prominently, our analysis is conditioned on the inductive biases of our architecture. Beyond the isotropy enforcement mentioned in Section~\ref{sec:isotropy}, there are many more constraints we set here, for example: 

\begin{itemize}
    \item The convolutional layer implicitly assumes translational invariance of the features
    \item Our summary statistic is always a weighted integral over the given NPCF which of course is a massive compression of the information
    \item By choosing MAFs as density estimators we are inherently biased toward a certain family of distributions for the posterior
\end{itemize}

Beyond the inductive biases we would like to point out additional caveats: As neural networks are always a non-convex optimization problem we can never find the global optimum which means that our findings should be seen as lower bounds only. We are furthermore tied to a finite training set which we optimize for. We have already shown in this work, that the available simulation budget has a large impact on the resulting constraints on cosmological parameters within our framework. 

While we therefore consider our architecture to be an excellent tool to approximate/bound the information content in NPCF, we ask the reader to keep in mind that this is subject to these conditions.

\section{Conclusions}
\label{sec:conclusions}

We have presented a novel framework for SBI that enables the efficient extraction of $N$-point information using learned, physically interpretable summary statistics with C3NN. Our approach combines the flexibility and performance of modern machine learning pipelines with a clear connection to traditional statistical descriptors of cosmological fields. By construction, each learned summary can be interpreted as a weighted sum of NPCFs at a given order, providing a transparent link between the network output and familiar correlation function–based analyses. In this sense, our method constitutes a physics-informed neural network, where strong architectural constraints encode prior physical knowledge directly into the model.

The interpretability is achieved through constraining design choices: the use of a single convolutional layer with rotationally symmetric filters, together with a recursive construction of higher-order moment maps. As a consequence, even architectural hyperparameters admit a physical interpretation. For example, the filter scale determines the largest spatial scales probed by the summaries, while the number of filters controls the resolution of scales that can be learned. These constraints substantially reduce the number of trainable parameters compared to standard CNN architectures.

In the limit of infinite training data, our architecture provides a principled framework to probe the information content of arbitrary NPCFs. In practice, using a modest set of 1250 simulations of simplified DES Y3-sized convergence maps, we are able to train models incorporating information up to the 4-point order. We find that the information content increases monotonically from 2nd to 4th order. While we do not observe a further gain from including the summary statistics of 5-point or even higher orders, we attribute this saturation to the limited simulation budget rather than an intrinsic lack of higher-order information in the simulated weak lensing convergence fields.

To explicitly verify that our model is sensitive to genuinely higher-order information, we performed inference on phase-reshuffled mock convergence maps. Phase reshuffling preserves the 2-point statistics information while erasing higher-order correlations. We find that, after reshuffling, the constraining power of all higher-order models collapse to that of the 2nd-order model. This provides a direct numerical confirmation of the analytic connection between our summaries and NPCFs, and demonstrates that the improved performance at higher order is indeed driven by information beyond the 2PCF.

We further compared our approach to a traditional 2PCF-based SBI analysis. In the baseline setup, we do not observe a statistically significant improvement over the 2PCF alone. However, we demonstrate that this limitation is purely driven by the small number of available simulations. We verify this conclusion in two complementary ways. First, we construct hybrid summaries in which the network is explicitly informed of the 2PCF and is trained only to learn additional information beyond it. Even with the original, limited simulation budget, these hybrid summaries outperform the 2PCF baseline. Secondly, we increase the training set by incorporating additional simulations not drawn from the prior. This necessitates a switch from neural posterior estimation (NPE) to neural likelihood estimation (NLE), allowing the prior to be changed after training. Doubling the number of training simulations in this manner leads to a substantial increase in constraining power. Together, these tests strongly support the conclusion that our method is capable of extracting additional information beyond the 2-point level, and that its ultimate performance is currently limited by the simulation budget rather than by the expressiveness of the architecture.

Our summary extraction framework opens several promising avenues for future work. One immediate direction is the interpretation of the learned filters themselves: after training, these filters may provide direct insight into the specific configurations and scales that are most informative at a given order. This could enable configuration-targeted inference tasks, such as searches for primordial non-Gaussianity in specific limits. On the methodological side, extending the architecture to other symmetry groups represents a natural next step. While a generalization to three-dimensional Euclidean data appears straightforward, a particularly compelling direction is the extension to non-Euclidean domains, such as the sphere. Since cosmological observations are naturally defined on $S^2$, this would allow for a more direct and natural application of our framework to real observational data.

\newpage

\begin{acknowledgements}
We would like to thank Annabelle Bohrdt, Anik Halder, Jed Homer, Tomasz Kacprzak, Yoann Launay, and Sankarshana Srinivasan for stimulating and insightful discussions. KL acknowledges support via the KISS consortium (05D23WM1) funded by the German Federal Ministry of Education and Research BMBF in the ErUM-Data action plan. We are also grateful for support from the Cambridge-LMU strategic partnership. DG was also supported by the European Union (ERC StG, LSS\_BeyondAverage, 101075919).

ZG and DG are deeply indebted to Stella Seitz for introducing and inspiring them to pursue research in the field of gravitational lensing, in which she made several foundational contributions. She recognized early on the potential of wide-area weak lensing surveys, advanced statistical tools, and machine learning methods, all of which are central to this work. Stella was deeply enthusiastic and actively involved in this project, but sadly did not live to see its final form. May she rest in peace, knowing that her legacy will live on through this and future works.
\end{acknowledgements}

\bibliography{bibliography}{}
\bibliographystyle{aasjournalv7}

\end{document}